\documentclass{egpubl}

\usepackage{eg2025}
\ConferencePaper
\authorCGFStandardLicense
\usepackage[T1]{fontenc}
\usepackage{dfadobe}  
\usepackage{cite}  
\biberVersion
\BibtexOrBiblatex
\electronicVersion
\PrintedOrElectronic

\usepackage{egweblnk}
\usepackage{amssymb}
\usepackage{booktabs}
\usepackage{graphicx}
\usepackage{enumitem}
\usepackage{soul}
\usepackage{dsfont}
\usepackage{gensymb}
\usepackage{microtype}
\usepackage{pifont}
\usepackage{threeparttable}
\usepackage{collect}
\usepackage{xspace}
\usepackage{xfrac}
\usepackage{array}
\usepackage{multirow}
\usepackage{wrapfig}
\usepackage{url}
\usepackage{amsmath} 
\usepackage{pifont}
\usepackage{xcolor}

\usepackage[normalem]{ulem}
\usepackage{algorithm}
\usepackage{algpseudocode}

\usepackage{xr}
\makeatletter
\newcommand*{\addFileDependency}[1]{
  \typeout{(#1)}
  \@addtofilelist{#1}
  \IfFileExists{#1}{}{\typeout{No file #1.}}
}
\makeatother

\DeclareGraphicsExtensions{.png,.jpg,.pdf,.ai}
\newcommand{\refFig}[1]{Fig.~\ref{fig:#1}}
\newcommand{\refTab}[1]{Tab.~\ref{tab:#1}}
\newcommand{\refSec}[1]{Sec.~\ref{sec:#1}}
\newcommand{\refAlg}[1]{Alg.~\ref{alg:#1}}
\newcommand{\refEq}[1]{Eq.~\ref{eq:#1}}
\newcommand{\method}[1]{#1}

\newcommand{\eg}{e.g.,\ }
\newcommand{\ie}{i.e.,\ }
\newcommand{\etal}{et~al.\ }

\newcommand{\mymath}[2]{
    \newcommand{#1}{\TextOrMath{$#2$\xspace}{#2}}
}
\mymath{\fractalcode}{\mathcal{F}}
\mymath{\function}{f}
\mymath{\probability}{p}
\mymath{\numfunctions}{N}
\mymath{\position}{\mathbf{x}}
\mymath{\attractor}{\mathcal{A}}
\mymath{\ifsmatrix}{M}
\mymath{\ifsbias}{\mathbf{b}}
\mymath{\inputimage}{{I_\textrm{ref}}}
\mymath{\pointset}{\mathcal{P}}
\mymath{\batchsize}{b}
\mymath{\trajectorylength}{l}
\mymath{\warmupiterations}{m}
\mymath{\rendering}{I}
\mymath{\params}{{\boldsymbol{\theta}}}
\mymath{\loss}{\mathcal{L}}
\mymath{\mipop}{\mathtt{mip}}
\mymath{\singval}{\sigma}
\mymath{\temperature}{\tau}
\mymath{\paramscurr}{{\params_\textrm{curr}}}
\mymath{\losscurr}{{\loss_\textrm{curr}}}
\mymath{\paramscand}{{\params_\textrm{cand}}}
\mymath{\losscand}{{\loss_\textrm{cand}}}
\mymath{\numpixels}{c}

\title{Learning Image Fractals\\\,Using Chaotic Differentiable Point Splatting}

\author[A. Djeacoumar \etal]
{
    \parbox{\textwidth}{\centering A. Djeacoumar\orcid{0009-0008-1919-450X}, F. Mujkanovic\orcid{0009-0009-9122-4408}, H.-P. Seidel\orcid{0000-0002-1343-8613}, T. Leimk\"uhler\orcid{0009-0006-7784-7957}}
    \\
    {\parbox{\textwidth}{\centering Max-Planck-Institut für Informatik, Saarbrücken, Germany}}
}

\begin{document}

\teaser{
   \vspace{-0.5cm}
    \includegraphics[width=0.99\linewidth]{figures/teaser_06_revised.ai}
    \centering
    \caption{
        We introduce a novel method to recover a fractal description from an image containing a self-similar shape. Our hybrid optimization achieves state-of-the-art fractal inversion, enabling the synthesis of intricate details at any desired scale -- illustrated here with 64x zoom-ins.
    }\label{fig:Teaser}
}
   
\maketitle
\begin{abstract}
Fractal geometry, defined by self-similar patterns across scales, is crucial for understanding natural structures. This work addresses the fractal inverse problem, which involves extracting fractal codes from images to explain these patterns and synthesize them at arbitrary finer scales. We introduce a novel algorithm that optimizes Iterated Function System parameters using a custom fractal generator combined with differentiable point splatting. By integrating both stochastic and gradient-based optimization techniques, our approach effectively navigates the complex energy landscapes typical of fractal inversion, ensuring robust performance and the ability to escape local minima. We demonstrate the method's effectiveness through comparisons with various fractal inversion techniques, highlighting its ability to recover high-quality fractal codes and perform extensive zoom-ins to reveal intricate patterns from just a single image.

\end{abstract}


\section{Introduction}
\label{sec:intro}

\begin{wrapfigure}{r}{0.17\textwidth}
    \vspace{-2mm}
    \begin{center}
        \includegraphics[width=0.17\textwidth]{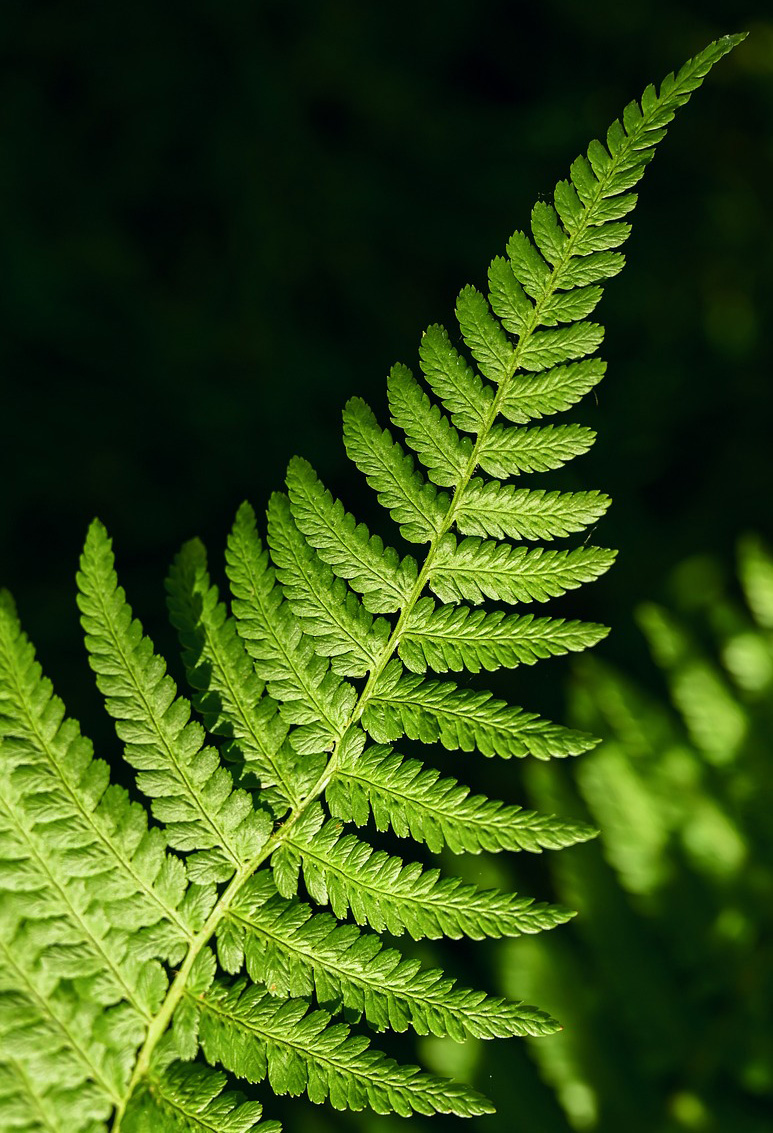}
    \end{center}
    \vspace{-1mm}
    \caption{A fern exhibiting self-similarities.}
    \vspace{-1mm}
    \label{fig:fern}
\end{wrapfigure}

The geometry of nature is often fractal~\cite{mandelbrot1982fractal}: 
Structures repeat themselves across different scales, forming self-similar patterns. 
A classic example is a fern, where each branch resembles both the overall fern structure and its smaller sub-branches (\refFig{fern}).
Similar observations can be made for coastlines, terrains, trees, river networks, blood vessels, etc.
Such intricate patterns can be explained by the repeated application of simple rules, referred to as \emph{fractal codes}.
Knowledge of underlying fractal codes is essential not only for describing and understanding complex systems~\cite{mandelbrot1982fractal,havlin1995fractals,turcotte1997fractals,song2005self,mandelbrot2013fractals} but also as an effective means of data compression~\cite{jacquinimage1992,fisher1994fractal}.
In this work, we develop a novel method for extracting fractal codes from an image containing a self-similar pattern using modern differentiable point splatting~\cite{kerbl20233d}. 

The \emph{fractal inverse problem} -- finding fractal codes that generate a given geometry -- is a long-standing formidable challenge~\cite{barnsley1988fractals} and typically framed as an optimization problem~\cite{Vrscay1991,lutton1995mixed,tu2023learning}. This presents two key challenges.

First, both the large-scale recursive application of fractal codes and the rendering of high-quality images are computationally demanding, especially within an iterative optimization framework. 
We observe that large-scale fractal synthesis, \ie recursively executing fractal codes hundreds of thousands of times per image, and well-designed rendering, including proper anti-aliasing, are both essential for achieving high-quality inversion results.
While computer graphics research has explored the efficient forward rendering of fractals over the past few decades~\cite{carpentercomputer1980,hart1991efficient,MARTYN2010167,daSilva2021}, there has been limited focus on improving efficiency in the inverse setting.
To address this gap, we introduce a highly efficient parallel fractal generator, paired with a custom differentiable renderer. 
Specifically, we propose a model that optimizes the parameters of an Iterated Function System (IFS)\cite{hutchinson1981fractals} to produce a point set that is differentiably rasterized\cite{zwicker2001surface,kerbl20233d}. 
Our model enables scaling up fractal synthesis to a quality level that facilitates state-of-the-art fractal inversion results.

The second key challenge in fractal inversion is the low dimensionality of the parameters, which results in highly non-convex objective energy landscapes that are notoriously difficult to navigate.
To address this challenge, stochastic optimization techniques, such as evolutionary algorithms~\cite{nettleton1994evolutionary,quirce2017cuckoo}, have been employed.
While these methods are effective at escaping local minima, they often follow inefficient optimization trajectories. 
In a different line of work, gradient-based optimization has been explored to solve the fractal inverse problem~\cite{melnik1998gradient,vrscay1999can,tu2023learning,bannister2024learnable}.
Gradients provide a strong guiding signal that enhances efficiency, but they are inherently prone to converging to local minima.
In this work, we make the observation that leveraging the synergies between stochastic and gradient-based techniques significantly outperforms all previous methods in solving the fractal inverse problem.
Our optimization employs an interleaved scheme that alternates between gradient descent~\cite{kingma2014adam} and simulated annealing~\cite{kirkpatrick1983optimization}, ensuring stable optimization trajectories while maintaining the ability to escape local minima.

Our approach recovers high-quality fractal codes from images. 
The key property of these codes is their ability to synthesize highly detailed patterns with virtually infinite resolution. 
These generative modeling capabilities allow for extreme zoom-ins on the original image, revealing intricate self-similarities and complex structures at any scale~(\refFig{Teaser}).
We conduct extensive comparisons with both classical and modern fractal inversion techniques, demonstrating the clear superiority of our method.

\noindent
In summary, our contributions are:
\begin{itemize}
    \item A novel, highly efficient image fractal inversion technique based on differentiable point splatting.
    \item An effective optimization routine that relies on synergies between stochastic and gradient-based optimization.
    \item State-of-the-art fractal inversion results alongside an in-depth analysis of the proposed algorithm and its components.
\end{itemize}


\section{Related Work}
\label{sec:related_work}


\subsection{Fractals}
\label{sec:fractals}

Fractals are complex geometric patterns that repeat infinitely at different scales, revealing self-similarity through recursive or iterative processes~\cite{mandelbrot1980fractal,mandelbrot1982fractal,peitgen2004chaos}.
Fractal geometry is not only crucial in many scientific disciplines~\cite{havlin1995fractals,turcotte1997fractals,song2005self,mandelbrot2013fractals}, but it has also been applied to the study of natural images~\cite{pentlandfractal1984, Turiel_2000} and visual art~\cite{redies2008fractal-like}, and is an indispensable tool for synthetic scene creation~\cite{demko1985construction,barnsley1988harnessing,ebert2002texturing,tzathas2024physically}.

A wide variety of algorithms for creating fractals and self-similarities exists. 
Grammar-based techniques use formal rules to iteratively generate complex structures~\cite{lindenmayer1968mathematical,rozenberg1980mathematical} with diverse applications~\cite{shlyakhter2001reconstructing,wonka2003instant,prusinkiewicz2012algorithmic}.
Noise-based approaches combine random functions at different scales~\cite{mandelbrot1968fractional,keshner19821}.
Escape-time algorithms generate fractals by iterating a function at each point until it escapes a boundary~\cite{mandelbrot2004fractals,julia1918memoire}.
Furthermore, self-similarity can be achieved by tiling~\cite{bandt1991self,fathauer2001fractal,ouyang2021self,chen2022learning}.
We build our approach on Iterated Function Systems, which generate fractals by repeatedly applying a set of geometric transformations~\cite{hutchinson1981fractals,barnsley1985iterated,barnsley1988fractals,elton1987ergodic}.
This method, which we review in \refSec{background}, is well-suited for point-based rendering~\cite{tu2023learning,bannister2024learnable} and has recently been shown to combine effectively with other fractal generation paradigms~\cite{schor2024into}.

Fractals have also been studied in 3D~\cite{hart1989ray,norton1982generation}, where efficient rendering is crucial~\cite{carpentercomputer1980,hart1991efficient,MARTYN2010167,daSilva2021}. 
Similarly, our optimization scheme relies on efficient synthesis and rendering to produce high-quality fractal images -- a property we found essential for achieving state-of-the-art inversions.

Our approach incorporates a deep differentiable pipeline. 
Fractals have been integrated into deep learning in various ways, such as through self-similar architectures~\cite{li2019lightweight}, reframing computations as neural networks~\cite{stark1991iterated}, deep equilibrium models~\cite{bai2019deep}, and training data augmentation~\cite{kataoka2020pre,anderson2022improving}.


\subsection{Fractal Inversion and Procedural Modeling}
\label{sec:fractal_inversion}

Obtaining fractal codes from an image is a notably intricate process~\cite{barnsley1986solution,barnsley1988fractals,Vrscay1991}. 
Since direct methods are only feasible for simple synthetic cases~\cite{berkner1997wavelet,struzik1994solution}, the task is generally formulated as an optimization problem.
Stochastic techniques have been explored to address the highly non-convex energy landscapes involved.
These include evolutionary algorithms optimizing the fractal code directly~\cite{jacques1993optimization,nettleton1994evolutionary,lankhorst1995iterated,gutierrez2000hybrid}, a partitioning of the fractal~\cite{sarafopoulos2006resolution}, or a polar reparameterization~\cite{collet2000polar}.
Similarly, swarm intelligence algorithms solve the problem by modeling stochastic natural processes~\cite{quirce2017cuckoo,galvez2018modified,galvez2021modified}.
Additionally, quadratic programming~\cite{ForteSolving1994} and expectation maximization~\cite{bloem2017expectation} have been applied to solve this task.

In a separate line of research, gradient-based approaches have been explored~\cite{vrscay1999can}. 
Differentiability has been achieved by using point set distances~\cite{melnik1998gradient}, moments~\cite{vrscay1989iterated,rinaldo1994inverse}, signed distance functions~\cite{kim2015quaternion}, and, recently, differentiable point splatting~\cite{tu2023learning,bannister2024learnable,scott2024differentiable}. Additionally, experiments have been conducted on directly regressing fractal codes using neural networks~\cite{grahamapplying2021}.
The approach most similar to ours is that of Tu~\etal~\cite{tu2023learning}. However, unlike their focus on representing low-resolution images with points arising from an Iterated Function System, we aim to recover infinite-resolution fractal structures.
To achieve this, we utilize high-quality renderings produced by an efficient forward model, along with a hybrid optimization framework to prevent getting stuck in local minima.

In addition to applications in data compression~\cite{jacquinimage1992,barnsley1993fractal, fisher1994fractal} and symmetry detection~\cite{liu2010computational,lukavc2017nautilus}, discovering self-similarities has also been studied as a powerful tool for generative modeling~\cite{poli2022self,karnewar3ingan2022,zhang2023rose,merrell2023example}.
Inverse procedural generative models have been studied in various forms, including tiling~\cite{vanhoey2013fly}, noise models~\cite{gilet2014local,heitz2018high}, handcrafted spatial patterns~\cite{LP-gi2000}, handcrafted material-specific procedures~\cite{guo2020bayesian}, and both fixed~\cite{hu2019novel,shi2020match} and learnable~\cite{hu2022inverse} procedural node graphs. However, to our knowledge, little to no inverse procedural models generate details at infinite scales, thus most of them lose detail when magnifying way past the scale of the reference image.
In contrast, in our work we learn a generative model that automatically extrapolates the target structure from a single scale to infinitely finer scales, constituting a significant advantage for infinite-resolution image synthesis.


\subsection{Multiscale Image Representation and Reconstruction}
\label{sec:multiscale}

Representing, synthesizing, and reconstructing images at multiple scales is a fundamental task in computer graphics and computer vision. 

Coarser representations of an image can be formalized using linear scale-space theory~\cite{iijima1959basic, witkin1987scale} and efficiently implemented using pyramids~\cite{burt1981fast, williams1983pyramidal}. 
In recent years, neural multiscale representations have gained significant attention~\cite{fathony2020multiplicative, chen2021learning, paz2022multiresolution, lindell2022bacon, saragadam2022miner, mueller2022instant, Belhe2023Discontinuity, mujkanovic2024ngssf}.
In contrast to our approach, all these approaches assume that a maximum resolution exists and a corresponding image is available.
However, we leverage image pyramids as part of our multiscale supervision structure.

The reverse task, inferring finer-scale images from coarser ones, is an active area of research in the super-resolution community.
For an overview, we refer to a recent survey~\cite{superresSurvey2023}.
Similar to our setting, the internal statistics of an image in the form of recurring patches has been studied~\cite{shechtman2007matching,zontak2011internal} to increase resolution~\cite{glasner2009super,ZSSR2018,bell2019blind}, remove blur~\cite{michaeli2014blind}, or synthesize new layouts~\cite{shaham2019singan,shocher2019ingan,zhou2018non}.
Unlike these methods, we model self-similarity using analytic functions.
This results in a narrower application domain, but enables image synthesis without scale limits.

Another approach to obtaining a multiscale representation is to fuse images from different scales. 
This is often done with an understanding of how the source images are related to each other~\cite{klashed2010uniview, halladjian2019scale, mohammed2017abstractocyte, tao2019kyrix, licorish2021adaptive}, but unstructured image collections have also been explored~\cite{wolski2024scalespacegan}.
In contrast, our approach takes a single image as input and infers generative rules to synthesize infinite-resolution details.


\subsection{Point Splatting}
\label{sec:point_splatting}

Point-based rendering has a long and rich history~\cite{gross2011point}. 
Early work focused on efficient hardware-accelerated point samples~\cite{grossman1998point}, while advancements in ``splatting'' reconstruction kernels, such as EWA~\cite{zwicker2001surface}, enabled high-quality image synthesis free from aliasing. 
The emergence of differentiable visual computing~\cite{li2019differentiable,spielberg2023differentiable} has further advanced this field, with soft reconstruction kernels facilitating the solution of inverse problems through differentiable point splatting~\cite{wiles2020synsin, yifan2019differentiable, lassner2021pulsar}. 
A notably efficient implementation of this approach is 3D Gaussian Splatting (3DGS)~\cite{kerbl20233d}, which marks the current state of the art in scene reconstruction from images. 
We adapt their splatting framework to recover 2D fractals and pair it with a stochastic optimization routine.

Similar to our work, 3DGS-based reconstruction has been treated as a stochastic process~\cite{kheradmand20243d}, akin to Stochastic Gradient Langevin Dynamics~\cite{brosse2018promises}. 
However, we demonstrate that mere gradient perturbations are insufficient to obtain high-quality fractal inversions.


\section{Background}
\label{sec:background}

\begin{figure}
    \includegraphics[width=\linewidth]{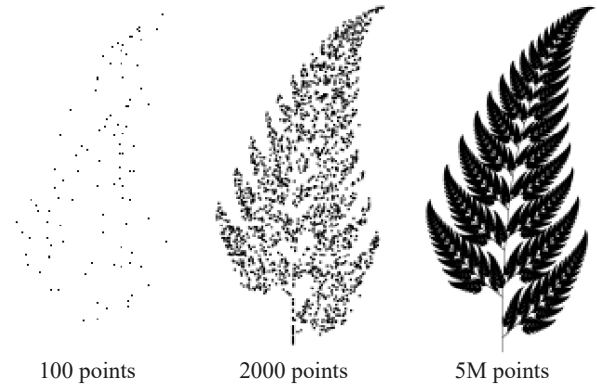}
    \caption{
    An IFS fractal generated using the chaos game (\refEq{chaos_game}) with varying numbers of points.
    }
    \label{fig:chaos_game}
\end{figure}

Here, we briefly review the concept of Iterated Function Systems (IFS)~\cite{hutchinson1981fractals,barnsley1985iterated} that implicitly define fractal geometry through the repeated application of geometric transformations.
Since this work focuses on \emph{image} fractals, we restrict our exposition to the 2D domain.
For a comprehensive introduction, we refer to Barnsley~\cite{barnsley1988fractals}.

An IFS fractal code
$
\fractalcode = 
\left\{ 
\left(  \function_i, \probability_i \right) 
\right\}_{i=1}^\numfunctions
$
is a set of \numfunctions continuous functions
$\function_i \in \mathds{R}^2 \rightarrow \mathds{R}^2$, each with an associated probability $\probability_i \in [ 0, 1 ]$, where
$\sum_i \probability_i = 1$.
Each function $\function_i$ deterministically moves a point at location \position to a new location $\function_i(\position)$ and is required to be contractive, \ie
\begin{equation}
\label{eq:contractive}
    \left\| \function_i(\position_1) - \function_i(\position_2) \right\|_2
    <
    \left\| \position_1 - \position_2 \right\|_2
    \quad
    \forall \position_1, \position_2 \in \mathds{R}^2.
\end{equation}
Based on this construction and starting from an arbitrary initial point $\position_0$, a sequence of points $\position_k$ can be generated using the ``chaos game''
\begin{equation}
\label{eq:chaos_game}
\setlength\fboxsep{7pt}
\boxed{
    \position_{k+1}
    =
    \function_{\pi_k}(\position_k),
}
\end{equation}
where the index $\pi_k$ is randomly selected from the set 
$\left\{i \right\}_{i=1}^\numfunctions$, 
with the probability of choosing index $i$ given by $\probability_i$.
It can be shown~\cite{hutchinson1981fractals,barnsley1985iterated} that the sequence $\position_k$ obtained through this Markov process converges to a compact set 
$\attractor \subset \mathds{R}^2$,
the \emph{attractor} of \fractalcode.
The attractor \attractor acts as a unique fixed point solely determined by the functions $\function_i$, \ie once a point $\position_k$ enters \attractor, subsequent applications of \fractalcode will cause the points to jump around within \attractor, but they will never leave it again.
Accumulating all points $\position_k$ effectively leads to an increasingly dense coverage of \attractor~(\refFig{chaos_game}).
The probabilities $\probability_i$ do not affect the shape of \attractor; they only influence the convergence rate of the chaos game by modulating local point densities and are typically derived from the determinants of $\function_i$~\cite{tu2023learning,anderson2022improving}.
Importantly, \attractor is a fractal by construction, as it consists of a union of (distorted) copies of itself~\cite{barnsley1985iterated}.
Henceforth, we will refer to sequences $\position_k$ as \emph{trajectories}.

The above exposition holds for any continuous and contractive functions 
$\function_i$~\cite{lutton1995mixed}.
However, in this work we follow standard practice~\cite{barnsley1986solution,gutierrez2000hybrid,grahamapplying2021,tu2023learning} and restrict ourselves to \emph{affine} functions of the form
\begin{equation}
\label{eq:affine_fct}
    \function_i(\position)
    =
    \ifsmatrix_i \position + \ifsbias_i,
\end{equation}
with $\ifsmatrix_i \in \mathds{R}^{2 \times 2}$
and $\ifsbias_i \in \mathds{R}^2$.

The fractal inverse problem can now be stated more precisely:
Given an input image $\inputimage$ depicting a (fractal) shape, find an IFS fractal code \fractalcode whose attractor \attractor matches that shape.
Once \fractalcode is obtained, the shape can be re-synthesized at virtually infinite resolution by rendering points generated through the chaos game.


\section{Method}
\label{sec:method}

\vspace{3mm}

\begin{quote}
    \textit{Chaos is a friend of mine.} \\
    \hspace*{\fill}-- Bob Dylan
\end{quote}
\vspace{2mm}

\begin{figure}
    \includegraphics[width=\linewidth]{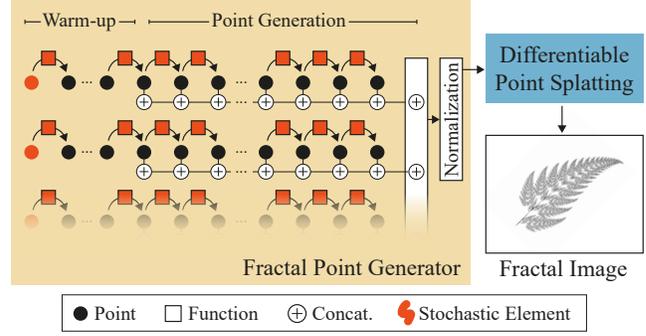}
    \caption{
    Overview of our model.
    The fractal point generator uses functions in $\fractalcode$ and employs a parallel stochastic scheme to efficiently and robustly execute the chaos game (\refEq{chaos_game}). The generated points are then rendered into an image using differentiable splatting.
    }
    \label{fig:overview}
\end{figure}

\noindent
The input to our algorithm is a binary or grayscale image \inputimage, representing a (fractal) shape of finite resolution or bandwidth, \eg a raster image.
Without loss of generality, we assume the shape is centered within \inputimage and padded by 25\% of the image's dimensions.
Our goal is to find an IFS code \fractalcode, such that when the chaos game from \refEq{chaos_game} is applied, the resulting attractor \attractor closely approximates the shape depicted in \inputimage.
Since \attractor is a fractal, we can use \fractalcode to render highly detailed, zoomed-in views at any scale, effectively surpassing the resolution limitations of \inputimage by an arbitrary amount.

We frame our task as an optimization problem over a set of affine functions 
$
\left\{
\function_i
\right\}_{i=1}^\numfunctions
$ 
(\refEq{affine_fct}), where the number \numfunctions of functions is set to a fixed conservative estimate.
As shown in \refFig{overview}, our model consists of two components. 
The first component is a fractal point generator, which runs the chaos game to generate a large point set 
$
\pointset
=
\left\{
\position_k
\right\}
$
that densely covers and thereby defines \attractor. 
To effectively incorporate this point generator into an optimization framework, we need to implement measures to improve efficiency and avoid overfitting to specific point trajectories, as detailed in \refSec{fractal_generator}.
The second component is a point splatting module, which differentiably rasterizes the points from the first component into an image. 
This process is described in \refSec{diff_point_splatting}.
Our model enables the formulation of an objective function in image space, which we minimize using a hybrid iterative optimizer that specifically addresses local minima via additional stochasticity, as outlined in \refSec{optimization}.


\subsection{Fractal Point Generator}
\label{sec:fractal_generator}

The purpose of this stage is the generation of points on the attractor \attractor via the chaos game.
We observed that achieving high-quality and stable results requires a large number of points, typically $| \pointset | \approx 500\textrm{k}$ in practice.
Therefore, evaluating \refEq{chaos_game} recursively for each optimization iteration is \emph{(i)} highly inefficient because of the lengthy computation sequence, and \emph{(ii)} unstable due to the need for gradients to flow through such an ultra-deep pipeline.

To address this issue, we use a generator that runs \batchsize trajectories of length \trajectorylength in parallel (rows in \refFig{overview}, left). 
Each trajectory is initialized with a uniformly sampled random point $\position_0$. Then, following \refEq{chaos_game}, we apply a sequence of transformations $\function_i$, randomly sampled from an uniform distribution to generate the point trajectories. Since $\position_0$ does not generally lie on \attractor, we treat the first \warmupiterations iterations as a warm-up phase, allowing the trajectory to converge onto \attractor. After this warm-up, we begin concatenating the generated points from all trajectories into a large buffer.

In each optimization iteration, our parallel pipeline samples \batchsize from the 
$\numfunctions^\trajectorylength$ possible combinations of $\function_i$, producing trajectories that start from randomly initialized positions $\position_0$.
This approach greatly helps the optimization focus on matching the attractor shape using the IFS rather than overfitting it using specific trajectories.
\refTab{sources_of_chaos} provides a list of all sources of stochasticity in our point generation pipeline, along with their respective sampling granularities.

\begin{table}
	\begin{center}
    	\caption
    	{
    		Sources of stochasticity in our framework, along with their granularities, \ie the levels at which random numbers are drawn.
    	}
        \label{tab:sources_of_chaos}	
		\begin{tabular}{lll}
            \toprule
			  Component & Source of Stochasticity & Granularity       \\
			\midrule
			    \multirow{2}{*}{Model} & Point Initialization & Trajectory \\
                & Function Selection & Point \\
            \midrule
                Optimizer & Simulated Annealing & Opt. Iteration \\
			\bottomrule
		\end{tabular}
	\end{center}
\end{table}

After generating all the points, we normalize them by applying a global uniform scaling and offset to center the point cloud and achieve the 25\% padding assumed for the shape in the input \inputimage.

\noindent
\textbf{Parameterization}
Our affine functions $\function_i$, as defined in \refEq{affine_fct}, must satisfy two key properties: 
\emph{(i)} They must be contractive, \ie \refEq{contractive} needs to hold, and 
\emph{(ii)} they should be appropriately constrained to prevent divergence during the highly stochastic optimization process.
To achieve this, we design a custom parameterization for the matrices $\ifsmatrix_i$ and offsets $\ifsbias_i$.

We parameterize $\ifsmatrix_i$ using the singular value decomposition (SVD)
$
    \ifsmatrix_i
    =
    U_i S_i V_i^T,
$
with $U_i, S_i, V_i \in \mathds{R}^{2 \times 2}$.
The diagonal matrix $S_i$ contains the non-negative singular values of $\ifsmatrix_i$. 
Each singular value is represented by a freely optimizable parameter to which a sigmoid function is applied, ensuring contractivity.
The orthonormal matrices $U_i$ and $V_i$ are represented by four optimizable parameters each, corresponding to their individual matrix entries.
After each optimization iteration, $U_i$ and $V_i$ are projected onto the nearest orthonormal matrices using the Gram-Schmidt process~\cite{cisse2017parseval}, ensuring that the decomposition remains a valid SVD.
Note that this represents an over-parameterization of $\ifsmatrix_i$ -- with 10 optimizable parameters for a $2 \times 2$ matrix. 
The offset $\ifsbias_i \in \mathds{R}^2$ is parameterized using a freely optimizable 2D vector, with the parameters passed through the hyperbolic tangent function.
This ensures that offsets remain within the range $[-1, 1]$.
We refer to the optimizable parameters of our model collectively as
$\params \in \mathds{R}^{12 \numfunctions}$.

In contrast to recent observations in previous work~\cite{bannister2024learnable,scott2024differentiable}, we discovered that our pipeline is insensitive to parameter initialization. 
Therefore, we use a straightforward uniform random initialization of \params.


\subsection{Differentiable Point Splatting}
\label{sec:diff_point_splatting}

To enable learning fractals from images, we differentiably rasterize the point set \pointset generated in the previous stage into an image \rendering. 
Each point is splatted as an isotropic 2D Gaussian~\cite{zwicker2001surface} using standard alpha blending~\cite{porter1984compositing}. 
For efficiency, we employ a tile-based rasterization pipeline~\cite{kerbl20233d}.

Splatting a stochastic approximation of an infinite-resolution fractal onto a finite-resolution image introduces challenges related to aliasing.
We address this issue with super-sampling.
Specifically, we splat the points into an image with a spatial resolution that is $5 \times 5$ higher than the input image \inputimage, using Gaussian splatting kernels with a standard deviation of 2.5 pixels. 
This is followed by averaging downsampling to the target resolution to arrive at our final output \rendering.
Through a parameter grid search, we found this combination of splat size and super-sampling factor to be an effective balance between detail preservation, anti-aliasing, gradient flow, and efficiency.


\subsection{Optimization}
\label{sec:optimization}

With our differentiable forward model established, we design an optimization algorithm to recover the fractal codes. 
Specifically, we optimize for the parameters \params of the IFS to generate an attractor image \rendering that closely matches the input image \inputimage.

\noindent
\textbf{Objective Function}
Our image-space objective function consists of four terms, addressing individual aspects of the problem setting:
\begin{equation}
\label{eq:full_loss}
    \loss =
    \lambda_\textrm{MSE} \loss_\textrm{MSE}
    + \lambda_\textrm{SSIM} \loss_\textrm{SSIM}
    + \lambda_\textrm{LPIPS} \loss_\textrm{LPIPS}
    + \lambda_\textrm{reg} \loss_\textrm{reg},
\end{equation}
where $\lambda_i \in \mathds{R}_+$ are corresponding weighting factors.
The first term,
\begin{equation}
\label{mse_loss}
    \loss_\textrm{MSE}
    =
    \sum_k
    \left\|
        \mipop_k
        \left(
            \rendering
        \right)
        -
        \mipop_k
        \left(
            \inputimage
        \right)
    \right\|_2^2,
\end{equation}
evaluates the reconstruction across a range of scales using image pyramids~\cite{burt1981fast, williams1983pyramidal}, where $\mipop_k$ denotes the $k$'th linear MIP map level.
The terms $\loss_\textrm{SSIM}$ and $\loss_\textrm{LPIPS}$ assess the error between \rendering and \inputimage using the D-SSIM~\cite{wang2004image} and LPIPS~\cite{zhang2018perceptual} metrics, respectively, promoting the alignment of overall shape structure.
Although the LPIPS metric has been specifically trained to evaluate \emph{natural} images, we found that $\loss_\textrm{LPIPS}$ significantly improves inversion results. 
This aligns with recent findings suggesting that pre-training neural networks with fractals is effective~\cite{kataoka2020pre,anderson2022improving}.

Finally, we found that an additional regularization in the form of
\begin{equation}
\label{eq:loss_reg}
    \loss_\textrm{reg}
    =
    \sum_{i=1}^\numfunctions
    \singval_{i, 1}^2
    +
    \singval_{i, 2}^2
    +
    \left\| \ifsbias_i \right\|_2^2
    +
    \lambda_\textrm{cond}   
    \left(
    \frac{\singval_{i, 1} + 1}{\singval_{i, 2} + 1}
    -1
    \right)^2
\end{equation}
is essential for stable optimization, where $\singval_{i, j}$ is the $j$-th singular value of $\ifsmatrix_i$ that can be read off directly from $S_i$.
While the singular values $\singval_{i,j}$ and offsets $\ifsbias_i$ are already constrained using saturating functions, we found that the optimization often drives them into the saturation region, leading to unfavorable dynamics. 
The first three summands of \refEq{loss_reg} address this issue by encouraging these values to remain as small as possible.
The last summand in \refEq{loss_reg} utilizes a stable approximation of the condition number of $\ifsmatrix_i$, helping to prevent singular matrices, which appear as streaks in the rendered fractal.

We set
$\lambda_\textrm{MSE} = 10$,
$\lambda_\textrm{SSIM} = 1$,
$\lambda_\textrm{LPIPS} = 2$,
$\lambda_\textrm{reg} = 10^{-2},$ and
$\lambda_\textrm{cond} = 10$
in all our experiments.

\noindent
\textbf{Hybrid Optimizer}
While our pipeline allows to compute 
$\nabla_\params \loss$
using automatic differentiation, we found that a pure gradient-based optimizer does not lead to satisfactory results.
This is due to the complex, recursive generation process, which is driven by a relatively small set of variables \params that all have a global impact.
As a solution, we employ a hybrid iterative optimization strategy that alternates between gradient descent and simulated annealing~\cite{kirkpatrick1983optimization}. 
The gradient descent component provides a stable path toward lower-energy states, while the simulated annealing component helps escape local minima.

Throughout the optimization, we maintain a ``temperature'' parameter \(\temperature\) that decreases linearly from 1 to 0.
Gradient descent is implemented using the Adam optimizer~\cite{kingma2014adam} with a fixed learning rate of $10^{-2}$ and default parameters otherwise. 
After every 250 iterations of gradient descent, we save the current parameters, denoted as \paramscurr, along with the current energy, denoted as \losscurr, and switch to a phase of simulated annealing.
In this phase, we successively sample 10 new parameter candidates 
$\paramscand = \paramscurr + \Delta\params$, 
where perturbations 
$\Delta\params \sim \mathcal{N}(\mathbf{0}, \sfrac{\temperature}{5})$
are drawn from an isotropic normal distribution. 
The corresponding energy \losscand is then evaluated. 
A candidate \paramscand is accepted as the new state \paramscurr with an acceptance probability
\begin{equation}
\label{eq:acceptance_prob}
    p(\paramscand)
    =
    \min
    \left(
    \exp
    \left(
        -10
        \cdot
        \frac
        {\losscand - \losscurr}
        {\temperature}
    \right), 1
    \right).
\end{equation}
This approach allows the optimization to occasionally move to a higher-energy state, helping to avoid getting trapped in local minima.
After each simulated annealing phase, we switch back to a gradient descent phase, resetting the momentum of the Adam optimizer.
This alternating scheme continues for the first half of the total optimization iterations, after which we stay in the gradient descent phase until completion.


\subsection{Implementation Details}
\label{sec:implementation}

We implemented our framework in PyTorch~\cite{paszke2017automatic}, using custom CUDA kernels for the point splatting module. Our source code, fractal data, and supplemental materials are available at \url{https://chaotic-fractals.mpi-inf.mpg.de}.

Our point generator creates point sets \pointset using $\batchsize = 2000$ trajectories in parallel, each with a length of $\trajectorylength = 250$ points. 
Using a warm-up phase of $\warmupiterations = 10$ iterations, this results in 480k points per optimization iteration. We found that this set of hyperparameters generates sufficient points for our optimization; using more points did not enhance inversion results.
For rendering, we adapted the differentiable rasterization pipeline from Kerbl~\etal~\cite{kerbl20233d} to enable efficient splatting of isotropic 2D kernels.
As our approach does not require any sorting of primitives, the asymptotic complexity of splatting reduces from 
$
\mathcal{O}
\left(
| \pointset | \cdot (\numpixels + \log | \pointset | )
\right)
$ 
to 
$
\mathcal{O}
\left(
| \pointset | \cdot \numpixels
\right)
$
compared to the original implementation, where \numpixels is the number of pixels in \rendering.
We use a resolution of $1024 \times 1024$ pixels for \inputimage and \rendering in all our experiments.
A full optimization requires 15k iterations.
We provide the pseudocode of our optimization algorithm in the Appendix~\ref{sec:appendix}.



\section{Evaluation}
\label{sec:Evaluation}

We evaluate our approach by comparing it against a wide range of baselines~(\refSec{comparisons}), presenting additional results~(\refSec{results}), and performing ablation studies~(\refSec{ablations}). 
Finally, we discuss the limitations of our method~(\refSec{limitations}).

By default, fractals for visualization and evaluation are generated with 50 million points per view and a supersampling rate of $8 \times 8$. To accelerate convergence, our point generator samples $\function_i$ based on probabilities proportional to their respective determinants \cite{anderson2022improving, tu2023learning}. Since differentiability is not necessary for evaluation, we use a basic hardware-accelerated point renderer implemented with PyOpenGL.
For certain combinations of zoomed-in views and IFS codes, achieving a point count of 50 million can be difficult or impossible, as it may be highly unlikely for points to fall within the small view window. In such cases, we limit the generation time to 20 minutes per view.
The corresponding images are labeled with a~$\blacktriangle$. \refTab{timing} presents a breakdown of the synthesis time for our fractal images across two GPU models.

\begin{table}
    \setlength{\tabcolsep}{7pt}
    \begin{center} 
        \caption
        {
        Timing breakdown per iteration of our approach on two different GPU models. During optimization, we employ differentiable point splatting, whereas for evaluation, we use hardware-accelerated point rendering with 100\emph{x} more points.
        }
\label{tab:timing}
\begin{tabular}{lrrrr}
\toprule
\multirow{2}{*}{Component} & \multicolumn{2}{c}{Optim. (500k\,pts.)} & \multicolumn{2}{c}{Eval. (50M\,pts.)} \\
& A40 & RTX\,3090 & A40 & RTX\,3090 \\
\midrule
Generation & 99.5\,ms & 158.9\,ms & 4.7\,s & 5.2\,s \\
Rendering & 1.2\,ms & 1.4\,ms & 0.1\,s & 0.1\,s \\
Sim. ann. & 2.0\,ms & 2.7\,ms & -- & -- \\
Backward & 116.6\,ms & 128.2\,ms & -- & -- \\
\bottomrule
\end{tabular}
\end{center}
\end{table}


\subsection{Comparisons}
\label{sec:comparisons}

\noindent
\textbf{Setup}
For our evaluation, we use a test suite of 100 randomly generated IFS fractals, following Anderson~\etal~\cite{anderson2022improving}. 
In addition to randomly sampling the IFS parameters, we also randomly select the number of functions \numfunctions from the range $\left\{ 3,10 \right\}$.

\noindent
\textbf{Baselines}
We compare our approach against a wide spectrum of existing approaches for fractal inversion or high-resolution image synthesis.

First, we consider the recent method \method{Learning Fractals}~\cite{tu2023learning}. 
Like our approach, this baseline uses differentiable point rendering to optimize IFS parameters but focuses on low-resolution reconstruction within a purely gradient-based framework. 
Since their default settings -- 300 points and an image resolution of $32 \times 32$ -- produced unsatisfactory results, we also present results with an increased point count of 600 and a higher image resolution of $256 \times 256$. 
Since the original implementation is not specifically optimized for efficiency, we were unable to run experiments with higher-quality settings.

Second, we evaluate two stochastic optimization approaches. 
Evolutionary programming~\cite{nettleton1994evolutionary} optimizes a population of solutions by randomly perturbing them with Gaussian noise to create offspring, and then selecting the fittest through competitive survival. 
We carefully re-implemented this method.
A more recent Cuckoo search approach~\cite{quirce2017cuckoo}, on the other hand, uses Lévy flight random walks and employs reproductive rules inspired by the egg-laying behavior of cuckoos. 
For this baseline, we adopted the code provided by the original authors.

Further, we present best-effort results for directly regressing IFS parameters from images using a neural network. 
In contrast to the approach in the seminal work~\cite{grahamapplying2021}, we found that a residual design -- regressing deviations from identity transformations -- yielded more stable results.
Training of the convolutional network is supervised on 8000 random fractals with IFS codes sorted according to their singular values.

Finally, to further contextualize the capabilities of our approach, we present results from a recent single-image super-resolution method. 
Specifically, we evaluate SwinIR~\cite{liang2021swinir}, a neural super-resolution architecture that integrates the spatially invariant filters of convolutional layers with Swin transformer~\cite{liu2021swin} layers, enabling the model to learn long-range dependencies across the image.
We train this model on 10k fractal images to upsample $1000 \times 1000$ images to a resolution of $8000 \times 8000$ pixels, following the setup used by the original authors.

For all IFS-based baselines, we assume a conservative number of functions, with $\numfunctions = 10$. 
To ensure fair comparisons, we utilized tuned hyperparameters for all baselines in cases where they produced better results.

\noindent
\textbf{Metrics}
To evaluate reconstruction quality across different scales, we synthesize 6 randomly sampled patches per test fractal, along with the full fractal, using zoom-in factors of up to 8x. 
All numerical evaluations are based on this extended set of 700 images.
To assess shape fidelity, we use the F1 score and the mean intersection over union (IoU). Additionally, we report the standard image quality metrics PSNR, SSIM~\cite{wang2004image}, and LPIPS~\cite{zhang2018perceptual}. We observe consistent inversion quality across GPU models (Nvidia A40 and RTX\,3090), with only a slight increase in optimization time with the RTX\,3090 (\refTab{timing}).

\begin{figure*}
    \centering
    \includegraphics[width=.92\linewidth]{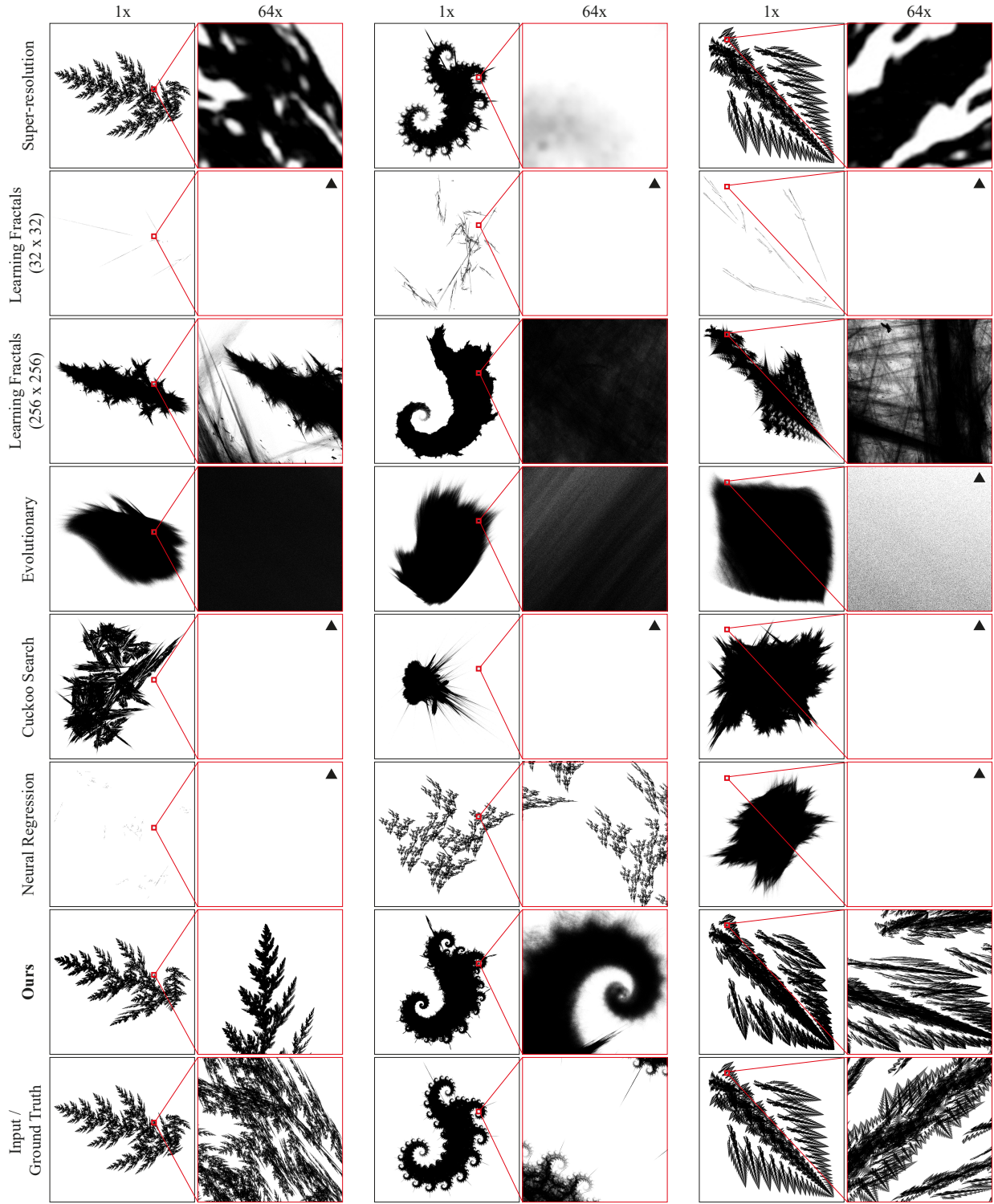}
    \caption{
    Qualitative fractal inversion results for various methods (rows). 
    For each instance (columns), the full fractal is shown on the left, with a 64x zoomed-in view on the right.
    The bottom row presents the input image \inputimage used for all methods (left) next to a zoomed-in view of the ground-truth fractal (right).
    Note that the zoomed-in views are consistent across rows.
    Images marked with a $\blacktriangle$ denote views that do not contain the full point count (see \refSec{Evaluation}).
    More visuals can be found in our supplemental materials.
    For continuous zoom-in visuals, please refer to our supplemental video.
    }
    \label{fig:qual_results}
\end{figure*}

\noindent
\textbf{Results}
We present the quantitative results of our analysis in \refTab{comparisons} and corresponding qualitative results in \refFig{qual_results}.
Please refer to our supplemental material for a complete gallery of qualitative results for the entire test set across methods.

Our approach significantly outperforms all baselines. 
The original version of \method{Learning Fractals} struggles to recover reasonable fractals due to its low-resolution $32 \times 32$ canvas. 
When we increase the resolution to $256 \times 256$ pixels -- the maximum supported by our modern hardware -- this baseline can recover coarse shapes of the target fractals but fails to capture finer structural details.
Despite our extensive efforts and several hours of optimization, stochastic methods using evolutionary programming and Cuckoo search did not converge to accurate results in our test setup. 
Similarly, neural network-based regression of IFS codes from images failed to produce coherent results, confirming previous findings~\cite{grahamapplying2021} and suggesting that generalizable solutions for the fractal inverse problem may not be feasible at this point. 
As expected, the super-resolution baseline performs well at coarser scales but ultimately produces structureless images when zoomed in due to its limited upsampling capacity.

\begin{table*}
	\centering
    \caption
    {
     Quantitative evaluation of our method compared to previous work. 
     We report shape-based (F1, IoU) and image-based (PSNR, SSIM, LPIPS) quality metrics alongside the time it takes to run each method on a single GPU.
     We include five fractal-based baselines and a pixel-based super-resolution method. 
     Unlike all other approaches, including ours, the super-resolution baseline is restricted to a maximum zoom-in capability of 8x.
    }
    \label{tab:comparisons}	
    \begin{threeparttable}	
		\begin{tabular}{lrrrrrr}
            \toprule
            Method & F1$\uparrow$ & IoU$\uparrow$ & PSNR$\uparrow$ & SSIM$\uparrow$ & LPIPS$\downarrow$ & Time \\
			\midrule
            \color{gray}\method{Super-resolution} \color{black}\cite{liang2021swinir}\tnote{*} & \color{gray}.917 & \color{gray}.854 & \color{gray}16.30 & \color{gray}.714 & \color{gray}.239 & \color{gray}12\,sec\\
            \midrule
            \method{Learning Fractals} ($32 \times 32$) \cite{tu2023learning} & .121 & .080 & 6.59 & .436 & .656 & 8\,min\\
            \method{Learning Fractals} ($256 \times 256$) \cite{tu2023learning} & .553 & .430 & 6.78 & .348 & .648 & 35\,min \\
            \method{Evolutionary} \cite{nettleton1994evolutionary} & .621 & .492 & 6.49 & .289 & .745 & 180\,min\\
            \method{Cuckoo Search} \cite{quirce2017cuckoo} & .490 & .351 & 3.43 & .140 & .825 & 540\,min\\
            \method{Neural Regression} \cite{grahamapplying2021}\tnote{*} & .161 & .106 & 5.33 & .358 & .709 & 16\,ms\\
            \textbf{Ours} & .691 & .572 & 8.91 & .449 & .428 & 55\,min\\
			\bottomrule
		\end{tabular}
    \begin{tablenotes}\footnotesize
        \item[*] Generalizable method that requires pre-training.
    \end{tablenotes}
    \end{threeparttable}
\end{table*}

\subsection{Additional Results}
\label{sec:results}

\begin{figure*}
    \includegraphics[width=\linewidth]{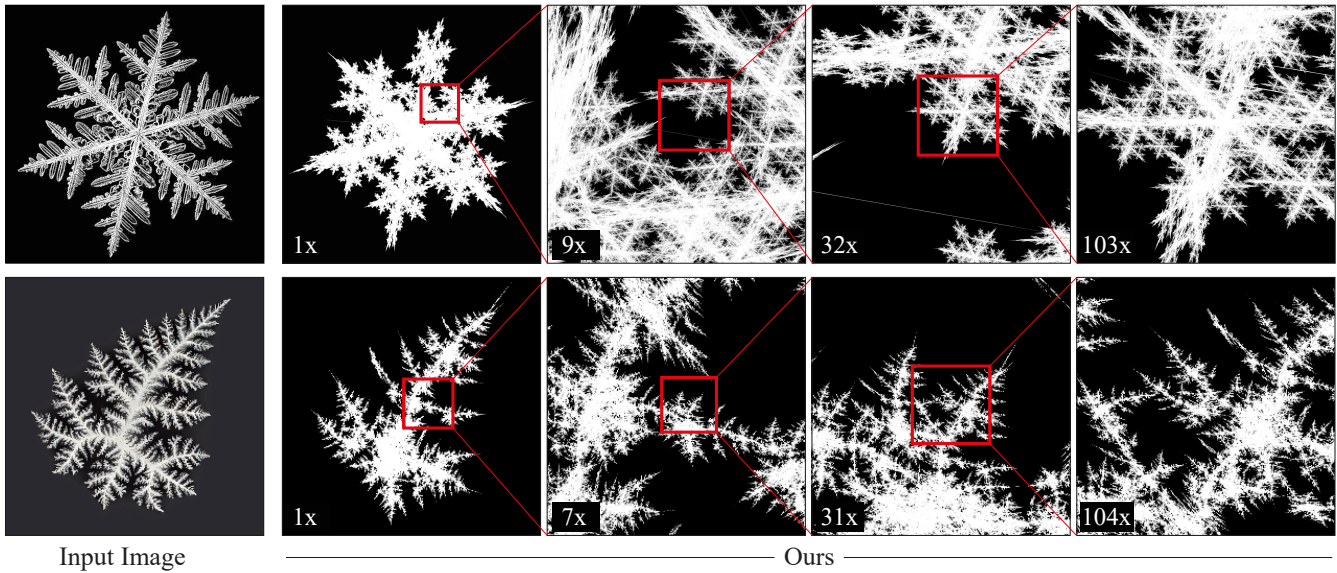}
    \caption{
    Inversion results for real images. Our representation enables infinite zoom-ins, continuously generating self-similar details at any scale while capturing intricate structural patterns. In contrast, natural structures exhibit self-similarity only within a limited scale range.
    }
    \label{fig:real}
\end{figure*}

\begin{figure}
    \includegraphics[width=\linewidth]{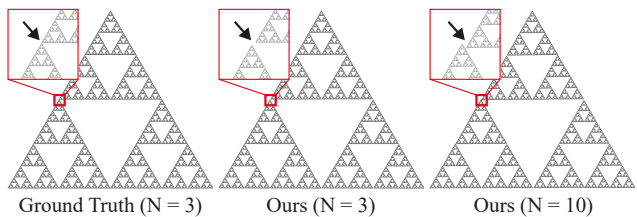}
    \caption{
    The Sierpinski triangle (left) alongside our reconstructions using different function counts \numfunctions (center and right). Our reconstructions closely match the ground truth, with only minor misalignments, visible in the insets, independent of \numfunctions.
    }
    \label{fig:sierpinski}
\end{figure}

In \refFig{real} we show fractal inversion results of real images. \refFig{sierpinski} presents our reconstruction of the famous Sierpinski triangle~\cite{sierpinski1915courbe} using two different values for the number of functions \numfunctions. 
The ground-truth fractal uses $\numfunctions = 3$. 
Our reconstructions closely match the ground truth, with only minor shifts, regardless of the \numfunctions value, provided it remains conservative. 
This happens because, when more functions are provided than needed, our pipeline converges to solutions where the extra functions are mapped to zero, resulting in negligible sampling probabilities.

\subsection{Ablations}
\label{sec:ablations}

\begin{figure*}
    \includegraphics[width=\linewidth]{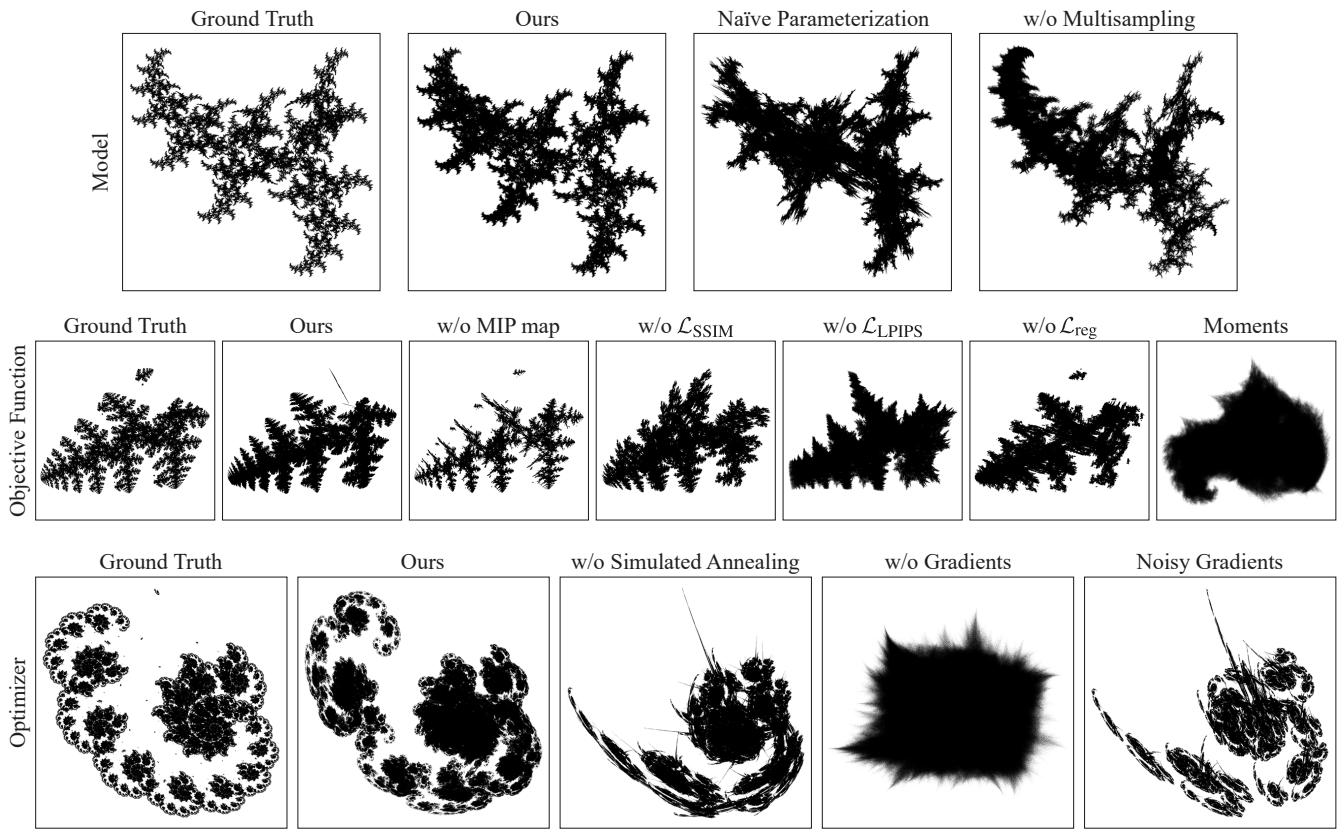}
    \caption{
    A qualitative overview of our ablational studies.
    We analyze different variations of our method in terms of the model (first row), the objective function (second row), and the optimizer (last row).
    }
    \label{fig:ablations}
\end{figure*}

Here, we analyze individual components of our method through ablation studies, examining variations in the model, objective function, and optimization routine. 
The results of the quantitative evaluations, based on 50 fractals, are presented in \refTab{ablations}, with corresponding qualitative examples shown in \refFig{ablations}.

\noindent
\textbf{Model}
We first explore an alternative to our SVD-based parameterization, where the affine functions $\function_i$ are parameterized directly by their matrix and vector entries (Naïve Parameterization). 
We found this approach to be prone to generating non-contractive functions, requiring us to significantly increase the regularization weight $\lambda_\textrm{reg}$ to obtain meaningful results.
Second, we omit our multisampling stage to investigate the importance of anti-aliasing in our framework (w/o Multisampling).
We observe that both components are important for high-quality inversions.

\noindent
\textbf{Objective function}
We omit individual components of our objective function to better understand their contributions to the final result. 
Specifically, we omit the MIP map in $\loss_\textrm{MSE}$, as well as the terms $\loss_\textrm{SSIM}$, $\loss_\textrm{LPIPS}$, and $\loss_\textrm{reg}$ individually.
Additionally, we replace our entire objective function with a classical alternative that minimizes the distance between moments~\cite{vrscay1989iterated,rinaldo1994inverse}.
Each term in our objective function plays a significant role in achieving our results, whereas the moments-based approach performs noticeably worse.

\noindent
\textbf{Optimizer}
To analyze our hybrid optimizer design, we evaluate our method using pure gradient-based optimization (w/o Simulated Annealing) and pure simulated annealing (w/o Gradients). 
Additionally, we explore a gradient-based approach in which we introduce carefully tuned Gaussian noise to the gradients in each optimization iteration to help escape local minima (Noisy Gradients).
Simulated annealing clearly aids in capturing the global shape, whereas a solution without gradients fails to converge to a meaningful result. 
Adding noise to the gradients provides only marginal improvement over a pure gradient-based solution.


\subsection{Limitations and Discussion}
\label{sec:limitations}
While our analysis shows that our approach marks the new state of the art in fractal inversion, it is important to acknowledge some limitations. 

Our results do not perfectly align with the ground truth; rather, the generated details only \emph{statistically} resemble the expected patterns. A single global solution exists and the extremely complex optimization landscape—characterized by numerous local minima and plateaus—makes perfect reconstruction of the ground truth challenging. At present, our hybrid optimization strategy seems best suited to effectively navigate this intricate space regardless of the initialization.

Our system, constrained by its analytical approach, \emph{always} results in perfectly self-similar structures. 
However, the self-similarity in natural images typically extends over only a few scales and is generally approximate due to the repeated influence of various physical forces acting on structures \cite{mandelbrot1982fractal}. For instance, the fern in \refFig{fern} exhibits about three scales of \emph{approximate} self-similarity, while the formation of snow crystals, a subject of study for decades \cite{libbrecht2005physics}, continues to present challenges. In such cases, our method may fail to find functions that represent the observed approximate self-similarity and hence inaccurately captures the ground truth (\refFig{real}). Using fractals, this discrepancy appears to be the trade-off for achieving arbitrary-scale image generation, as evidenced by our significant zoom-ins.

For future work, we envision a fractal system that incorporates more expressive functions as building blocks~\cite{lutton1995mixed} -- potentially parameterized by neural networks or enhanced by the injection of physical forces. Such an approach could enable better representation and simulation of complex self-similar geometries, while also achieving generative capabilities across scales.

A minor technical limitation of our approach is the fixed number of functions used in our model. A potential avenue for future work is the explicit optimization of sampling probabilities to remove this restriction, which likely requires an appropriate parameterization~\cite{jang2016categorical}.


\section{Conclusion}
\label{sec:conclusion}

We proposed a novel approach to solving the fractal inverse problem, which includes a differentiable forward model that implements the chaos game based on Iterated Function Systems and a point splatting module that renders the generated fractal point set into an image. 
Notably, we demonstrated that combining large-scale fractal generation with high-quality rendering, along with a hybrid of gradient-based and stochastic optimization, leads to state-of-the-art results in inverting image fractals.
This was validated through extensive comparisons with a diverse range of classical and modern methods.

We envision fractals as optimizable graphics primitives that, due to their point-based nature, can seamlessly integrate with recent advancements in Gaussian scene representations.
Fractals can serve as a fundamental tool for representing content across a wide range of scales -- capabilities that are challenging to achieve with classical (e.g., meshes) or neural representations (e.g., neural radiance fields~\cite{mildenhall2020nerf}).
To fully realize this potential, future research must extend robust fractal optimization frameworks to operate in 3D~\cite{norton1982generation,schor2024into} and incorporate appearance~\cite{bannister2024learnable}. 
These advancements will unlock new opportunities for fractals to contribute to the evolving landscape of scene representations and image synthesis.

\begin{table*}
    \setlength{\tabcolsep}{7pt}
	\begin{center}
    	\caption
    	{
    		Ablations.            
    	}
        \label{tab:ablations}	
		\begin{tabular}{llrrrrr}
            \toprule
            Component & Ablation & F1$\uparrow$ & IoU$\uparrow$ & PSNR$\uparrow$ & SSIM$\uparrow$ & LPIPS$\downarrow$  \\
			\midrule
            \multirow{2}{*}{Model} & Na\"{i}ve Parameterization & .583 & .467 & 9.12 & .481 & .431  \\
            & w/o Multisampling & .561 & .445 & 9.16 & .481 & .441 \\
            \midrule
            \multirow{5}{*}{Objective Fct.} & w/o MIP map & .363 & .286 & 9.08 & .525 & .512 \\
            & w/o $\loss_\textrm{SSIM}$ & .611 & .483 & 8.41 & .436 & .454 \\
            & w/o $\loss_\textrm{LPIPS}$  & .544 & .440 & 8.86 & .466 & .511 \\            
            & w/o $\loss_\textrm{reg}$ & .596 & .474 & 8.66 & .449 & .476 \\            
            & Moments & .542 & .408 & 4.17 & .170 & .818 \\

            \midrule
            \multirow{3}{*}{Optimizer} & w/o Simulated Annealing & .594 & .468 & 8.72 & .459 & .441 \\
            & w/o Gradients & .496 & .377 & 7.14 & .296 & .736 \\
            & Noisy Gradients & .578 & .455 & 8.93 & .471 & .442 \\
            \midrule
            \textbf{Ours} & & .611 & .488 & 9.23 & .482 & .415 \\
            
			\bottomrule
		\end{tabular}
	\end{center}
\end{table*}

\bibliographystyle{eg-alpha}
\bibliography{paper}

{
\vspace{50mm}
\appendix
\section{Pseudocode}
\label{sec:appendix}
We provide pseudocode of our optimization algorithm in \refAlg{pseudocode}.

\begin{algorithm}
\caption{Fractal Code Optimization\\
\small
$\inputimage$ : Input Image\\
$\batchsize$ : Batch Size of the Fractal Point Generator\\
$\params$ : Optimizable Fractal Parameters
}
\label{alg:pseudocode}
\begin{algorithmic}
\small
\State $\params \gets$ InitializeParameters() \Comment{Uniform Random}
\State $i \gets 0$ \Comment{Iteration Count}
\While{not converged}
    \State $\temperature \gets $ LinearAnneal($i$)\Comment{Temperature}
    \State $\pointset \gets$ \Call{GenerateFractalPoints}{$\params, \batchsize$}
    \State $\rendering \gets $ \Call{Render}{$\pointset$}
    \State $\loss \gets $\textit{Loss}($\rendering, \inputimage, \params$) \Comment{Loss : \refEq{full_loss}}
    \State $\params \gets $ ADAM($\nabla_\params\loss$)\Comment{Backprop \& Step}
    \If{DoHybridOptimization($i$)} \Comment{Hybrid Optimization}
        \State $\paramscurr \gets \params$
        \State $\losscurr \gets \textrm{MSE}(\rendering, \inputimage)$\Comment{Current Energy}    
        \For{$j \gets 0$ to $10$} 
            \State $\Delta\paramscand \gets \mathcal{N}(\mathbf{0}, \sfrac{\temperature}{5})$
            \State $\paramscand \gets \paramscurr + \Delta\paramscand$\Comment{Sample Neighbour}
            \State $\pointset \gets$ \Call{GenerateFractalPoints}{$\params, \batchsize$}
            \State $\rendering \gets $\Call{Render}{$\pointset$}
            \State $\losscand \gets \textrm{MSE}(\rendering, \inputimage)$ \Comment{Candidate Energy}
            \vspace{1mm}
            \State $p(\paramscand) \gets$ AcceptanceCriteria($\losscand, \losscurr, \temperature$) \Comment{\refEq{acceptance_prob}}
            \vspace{1mm}
            \If{$\losscand < \losscurr$ \textbf{or} $Random() < p(\paramscand)$}
                \State $\paramscurr \gets \paramscand$
                \State $\losscurr \gets \losscand$
            \EndIf
        \EndFor
        \State $\params \gets \paramscurr$
        \State Reset ADAM momentum
    \EndIf
\EndWhile
\vspace{2mm}
\hrule
\vspace{2mm}
\Function{GenerateFractalPoints}{$\params, \batchsize$}\Comment{\refSec{fractal_generator}}
    \State $\position_0 \gets$ SampleRandomPoints(b)
    \State $G \gets$ ConstructFractalPointGenerator($\params, \batchsize$)
    \State $\pointset' \gets G(\position_0)$
    \State $\pointset \gets$ RemoveInitPoints() \Comment{Warm-Up (\refSec{implementation})}
    \State \Return $\pointset$
\EndFunction
\vspace{2mm}
\Function{Render}{$\pointset$}\Comment{\refSec{diff_point_splatting}}
    \State $\rendering' \gets$ $\mathcal{R}(\pointset)$ \Comment{Differentiable Rasterization}
    \State $\rendering \gets$ SuperSample($\rendering'$) \Comment{Supersampling}
    \State \Return $\rendering$
\EndFunction

\end{algorithmic}
\end{algorithm}
}
\end{document}